\documentclass[reprint, twocolumn, secnumarabic, amsmath, amssymb, nobibnotes, aps, prd]{revtex4-1}

\usepackage{graphicx}
\usepackage{dcolumn}
\usepackage{bm}

\begin{document}

\preprint{APS/123-QED}

\title[Electrostatic energy and phonon properties of Yukawa crystals]{Electrostatic energy and phonon properties of Yukawa crystals}

\author{A. Kozhberov}
\email{kozhberov@gmail.com}

\affiliation{Ioffe Institute, Politekhnicheskaya 26, Saint Petersburg, 194021, Russia}%

\date{\today}

\begin{abstract}
We study electrostatic and phonon properties of Yukawa crystals. It is shown that in the harmonic approximation these systems which is in use in the theory dusty plasma can be described analytically by the model from the theory of neutron stars and white dwarfs. Using this approximation we consider properties of body-centred cubic (bcc), face-centred cubic (fcc), hexagonal close-packed (hcp), and MgB$_2$ lattices. MgB$_2$ and hcp lattices have never been studied earlier in the context of Yukawa systems. It is shown that they never possess the smallest potential energy and the phase diagram of stable Yukawa crystals contains bcc and fcc lattices only. However, corrections to the charge density $\propto (\kappa a)^4$ can noticeably change the structural diagram of Yukawa systems. The developed analytical model also allows to describe low-temperature effects where numerical simulations are difficult.

\begin{description}
\item[PACS numbers]
52.27.Gr, 52.27.Lw, 52.25.Kn, 05.70.Ce
\end{description}
\end{abstract}

\maketitle

\section{Introduction}
A Yukawa system is a system of point-like charged particles immersed in a neutralizing background. Usually it is assumed that all these particles are identical and have the electric charge $-Ze$ and mass $M$, where $e$ is the electron charge absolute value. The background in the Yukawa systems is non-uniform and can be described by the inverse screening length $\kappa$. For instance, if the background consists of electrons, $\kappa\equiv \sqrt{4 \pi e^2 \partial n_e / \partial \mu_e}$, where $n_e$ and $\mu_e$ are electron mean number density and chemical potential, respectively. Systems with a uniform background are called a Coulomb systems. These systems are widely used in various branches of physics.

It is believed that the matter in degenerate stars, namely in white dwarf cores and neutron star crusts at not too low densities, consists of atomic nuclei (ions) immersed into neutralizing background of electrons \cite{ST}. As a star cools, ions crystallize (e.g., \cite{HPY07,HB09}). According to Ref. \cite{W09}, this crystallization process has been inferred from observations of oscillations of a white dwarf. While observations of transients indicate that the neutron star crust is solid \cite{Sh1,Sh2,Sh3}. In degenerate stars electrons are mostly strongly degenerate and can be characterized by the Thomas-Fermi (TF) wavevector $\kappa_{\rm TF}\approx0.185Z^{1/3}(1+x_{r}^2)^{1/4}/x_{r}^{1/2}$, where $x_{r}\equiv p_{F}/(m_e c)$ is the electron relativity parameter, $p_{F}$ is the electron Fermi momentum and $m_e$ is the electron mass (e.g., \cite{Ba02}). Notice that the ordered structures can form in red giants and brown dwarfs \cite{Fort2}.

Yukawa systems are also widely used in the dusty plasma physics (e.g., \cite{Fort1}). In the simplest model the dusty plasmas consists of charged dust grains and the background which is formed by nondegenerate ions and electrons (here we follow the model which was developed in \cite{FH93,FH94I}). For nondegenerate electrons $\partial n_e/\partial\mu_e=n_e/(k_{B}T)$ and similarly for ions. Hence, in this situation $\kappa$ is the inverse Debye (D) length: $\kappa_{D}^2=\kappa_{De}^2+\kappa_{Di}^2\equiv4\pi e^2n_e/(k_{B}T)+4\pi e^2Z_i^2n_i/(k_{B}T)$, where $Z_i$ and $n_i$ are the charge number and mean number density of background ions, $T$ is the temperature of the system (the temperature of background ions is taken equal to the temperature of dust particles) and $k_{B}$ is the Boltzmann constant.

Both such different systems can be described by a simple model called Yukawa crystal: point-like charged particles arranged into a lattice and the neutralizing background characterized by parameter $\kappa$. In this paper for the first time we show that results of molecular dynamic simulations of Yukawa crystals \cite{FH94II,FHD97} can be verified by the theoretical model which was developed in \cite{Ba02} for the degenerate stars (in \cite{Ba02} they are called ``Coulomb crystals with polarizable background'').

In \cite{Ba02} and \cite{FHD97} only the body-centred cubic (bcc) and face-centred cubic (fcc) lattices were considered. As well as in some other theoretical parers (e.g., \cite{HD03,HR04,KM12,VK15}) which by different approachers qualitatively or/and quantitatively prove results of simulations from \cite{FHD97}.  The 3D hexagonal Yukawa crystals were never studied by molecular-dynamic simulations as it was done in \cite{FHD97} for the bcc and fcc lattices. On the other hand, it is known that the ground state of 2D crystals has the hexagonal symmetry (e.g., \cite{BL08}). Also the laboratory experiments (in many experiments dusty plasma are not a bulk 3D system because of the presence of different forces such as gravity, ``shadow forces'', thermal forces and some others and can not be described by the model of Yukawa crystal) show that the strongly coupled dusty plasma forms a complicated crystal structure (e.g., \cite{TM94,CI94,TM95,M96,ZI00,FKM04,Fort1}). Analysis of experiments carried out onboard International Space Station under microgravity conditions shows that plasma crystals have a structure which contains numerous bcc, fcc, and hexagonal close-packed (hcp) clusters with the prevailing contribution of the latter two \cite{K09,K10}. The similar results give different computer simulations (e.g., \cite{DD07,RS15,BHB10}). Hence a formation of the hexagonal Yukawa crystals is quite probable.

Despite the interest to the dusty systems the structural diagram of Yukawa crystals have received relatively little attention. The transition between the bcc and fcc lattices was firstly obtained from molecular-dynamic simulations in Ref. \cite{RKG88}, developed in Ref. \cite{FHD97} and later was only proved in a few papers (e.g., \cite{HR04}). Transitions between other lattices have not been ever considered and the structural diagram has not been studied analytically. In the present work we study properties of the hcp Yukawa crystal. In the harmonic lattice approximation we calculate its total potential energy as a sum of electrostatic energy, energy of zero-point vibration, and phonon free energy. This total potential energy is used to study the structural transitions between the hcp and cubic (bcc and fcc) lattices in dusty plasmas. One of the advantages of the harmonic lattice approximation is ability to take into account the low-temperature effects where numerical simulations are difficult to provide. The importance of high-order corrections to the charge density is also discussed.

\section{Electrostatic energy}
Yukawa crystals were investigated in \cite{FH94I} via molecular dynamics simulations in a cubical domain with the side length $L$ and periodic boundary conditions. Its volume is $V\equiv L^3$ and $N$ is the number of charged point-like particles in the crystal. The total potential energy $U$ of such systems is given by (equation (29) from \cite{FH94I})
\begin{eqnarray}
&&U=N\frac{Z^2e^2}{2} \label{U}\\
&&\times\left[\sum_{j'\neq j}\Phi({\bf r}_j-{\bf r}_{j'})-\frac{4\pi n}{\kappa^2_{D}}-\kappa_{D}+\sum_{m\neq0}\frac{e^{-\kappa_{D}mL}}{mL}\right], \nonumber
\end{eqnarray}
where
\begin{equation}
\Phi({\bf r})=\sum_{m\neq0}\frac{\exp(-\kappa_{D} |{\bf r}-{\bf m}L|)}{|{\bf r}-{\bf m}L|},
\end{equation}
${\bf m}=(m_1,m_2,m_3)$ denotes integer triplet, $n\equiv N/V$ is the mean number density of charged point-like particles. According to the electroneutrality condition, $Zn=Z_in_i-n_e$. The position of an $j$th particle in the crystal is given by radius vector ${\bf r}_{j} = {\bf X}_j + {\bf u}_j$, where ${\bf X}_j$ is the particle equilibrium position and ${\bf u}_j$ is the displacement. Equation (\ref{U}) was obtained from the Poisson's equation
\begin{equation}
\Delta \Psi({\bf r})=-4\pi \rho ({\bf r}), \label{Puas}
\end{equation}
where the charge density
\begin{equation}
\rho({\bf r})=-Z\sum_{j}\delta({\bf r}-{\bf r}_j)+Zn-\frac{\kappa_{D}^2}{4\pi} \left(\Psi({\bf r})-\overline{\Psi}\right), \label{pho}
\end{equation}
where
\begin{equation}
\overline{\Psi}\equiv \frac{1}{V}\int_V\Psi({\bf r})d{\bf r}.
\end{equation}
In this approach, the variation of the potential $\Psi({\bf r})$ over $V$ should be much smaller than the thermal energy. The next order correction to the charge density is proportional $\kappa_{D}^4$.

If all particles are fixed in their equilibrium positions, $U$ reduces to the electrostatic (Madelung) energy $U_{M}$. For a lattice with $N_{\rm cell}$ particles in the elementary cell, equilibrium positions are given by $\textbf{X}_j=\textbf{X}_{lp}=\textbf{R}_{l}+\boldsymbol{\chi}_{p}$, where $\textbf{R}_l$ is the lattice vector, $\boldsymbol{\chi}_{p}$ is the basis vector of the $p$-particle ($p=1...N_{\textrm{cell}}$) in the elementary cell and $l=(l_1,l_2,l_3)$ is the integer triplet. The reciprocal lattice is formed by vectors $\textbf{G}_{b}$, where $\textbf{b}=(b_1,b_2,b_3)$ is the integer triplet.

$U_{M}$ can be found analytically at fixed $n$ and $L \rightarrow \infty$. Using the Ewald transformation, it is possible to derive a rapidly converging expression for the Madelung energy of the Yukawa lattice \cite{Ba02}:
\begin{eqnarray}
\frac{U_{M}}{NZ^2 e^2}&\equiv&\frac{\zeta}{a}=\frac{1}{N_{\rm cell}}\sum\limits_{l,p,p'} \left(1-\delta_{l0}\delta_{pp'}\right) \frac{E_{-}+E_{+}}{4Y_l} \nonumber \\
&+&\frac{1}{N_{\rm cell}^2}\sum\limits_{b,p,p'} \frac{2\pi n}{G_b^2+\kappa^2}e^{-\frac{G_b^2+\kappa^2}{4A^2}}e^{-i{\bf G}_b(\boldsymbol{\chi}_p-\boldsymbol{\chi}_{p'})}\nonumber \\
&-&\frac{\kappa}{2}{\rm erf}\left(\frac{\kappa^2}{2A}\right)-\frac{A}{\sqrt{\pi}}e^{-\frac{\kappa^2}{4A^2}}-\frac{2\pi n}{\kappa^2}, \label{UTF}
\end{eqnarray}
where $E_{\pm}=e^{\pm\kappa Y_l}\left[1-{\rm erf}\left(AY_l\pm \kappa/(2A)\right)\right]$, ${\rm erf}(z)$ is the error function, ${\bf Y}_l={\bf R}_l+\boldsymbol{\chi}_p-\boldsymbol{\chi}_{p'}$, and $a \equiv (4\pi n/3)^{-1/3}$ is the Wigner-Seitz radius. Parameter $A$ is chosen in such way that the summation over direct and reciprocal lattice vectors converge equally rapidly. For lattices in consideration, $Aa\approx2$. Parameter $\zeta$ is called the Madelung constant. It depends on the type of the lattice and $\kappa a$. Equation (\ref{UTF}) for $U_{M}$ has the same form for $\kappa_{D}$ and $\kappa_{\rm TF}$. For this reason the subscript in $\kappa$ is omitted.
\begin{table*}
\caption{\label{tab:1}  The Madelung constants of the bcc, fcc, hcp, and MgB$_2$ lattices.}
\begin{ruledtabular}
\begin{tabular}{cccccccccc}
$\kappa a$ & \multicolumn{2}{c} {bcc lattice} & \multicolumn{2}{c} {fcc lattice} &\multicolumn{3}{c} {hcp lattice}  &\multicolumn{2}{c} {MgB$_2$ lattice}  \\
\cline{2-10}
& $-\zeta_{\rm HF}$ & $-\zeta$ & $-\zeta_{\rm HF}$ & $-\zeta$ & $h_{\min}/a_{\rm lat}$ & $-\zeta_{\min}$ & $-\zeta$ & $h_{\min}/a_{\rm lat}$ & $-\zeta$ \\
\hline
0.0 & $0.895929$ & $0.895929256$ & $0.895873$ & $0.895873616$ & $1.635639$ & $0.895838451$ & $0.895838120$ & $0.593936$ & $0.894505630$ \\
0.2 & $0.900074$ & $0.900073612$ & $0.900020$ & $0.900020482$ & $1.635630$ & $0.899985873$ & $0.899985549$ & $0.593958$ & $0.898663839$ \\
0.5 & $0.921671$ & $0.921671339$ & $0.921631$ & $0.921630646$ & $1.635543$ & $0.921598798$ & $0.921598509$ & $0.594074$ & $0.920331769$ \\
1.0 & $0.996706$ & $0.996706468$ & $0.996701$ & $0.996701309$ & $1.635278$ & $0.996677534$ & $0.996677339$ & $0.594458$ & $0.995586492$ \\
2.0 & $1.269026$ & $1.269025941$ & $1.269079$ & $1.269079142$ & $1.634495$ & $1.269071279$ & $1.269071235$ & $0.595636$ & $1.268452721$ \\
3.0 & $1.651144$ & $1.651143676$ & $1.651194$ & $1.651193657$ & $1.633786$ & $1.651192170$ & $1.651192165$ & $0.596801$ & $1.650930425$ \\
4.0 & $2.091283$ & $2.091283389$ & $2.091309$ & $2.091308661$ & $1.633349$ & $2.091308471$ & $2.091308471$ & $0.597578$ & $2.091219086$ \\
\end{tabular}
\end{ruledtabular}
\end{table*}

For lattices with $N_{\rm cell}=1$ equation (\ref{UTF}) was derived in \cite{Ba02} and was used for the bcc and fcc lattices. For lattices with $N_{\rm cell}>1$, the expression for $U_{M}$ is given here for the first time. In \cite{FH94II} electrostatic energy was calculated from the molecular dynamics simulations as a limiting value of $U$ at $T\rightarrow0$. Madelung constants for the bcc lattice obtained in \cite{FH94II} are given in Tab. \ref{tab:1} as $\zeta_{\rm HF}$. Our results based on Eq. (\ref{UTF}) are given as $\zeta$ in Tab. \ref{tab:1}. For the bcc lattice they coincide with the results of \cite{FH94II} and \cite{RKG88}, but more significant digits are given. For the fcc lattice both calculations are also consistent. Equation (\ref{UTF}) allows to calculate the Madelung energy for any lattice much more accurate and fast than molecular dynamics simulations.

Usually (e.g., \cite{BPY01}), for the hcp lattice the distance between hexagonal layers $h$ in the elementary cell is assumed to be $h_0\equiv\sqrt{8/3}a_{\rm lat}\approx1.632993a_{\rm lat}$, where $a_{\rm lat}$ is the lattice constant. This value comes from the problem of close-packing of equal spheres. However in Yukawa crystals charged particles are point-like and it is not obvious that the electrostatic energy of the hcp lattice achieves minimum at $h=h_0$ in this case. It is more correct to consider the hcp lattice with $h=h_{\min}$ that corresponds to the minimum of $U_{M}$. $h_{\min}$ values obtained for several $\kappa a$ are also presented in Tab. \ref{tab:1}.  At $\kappa a=0$ (uniform background) the Madelung energy of the hcp lattice reaches minimum at $h_{\min}\approx1.635639a_{\rm lat}>h_0$. The similar situation is found at $0<\kappa a<5$: $h_{\min}/a_{\rm lat}$ is always slightly greater than $\sqrt{8/3}$ and decreases when $\kappa a$ increases. This small difference does not lead to significant change of $U_{M}$. In Tab. \ref{tab:1}, $\zeta_{\min}$ corresponds to the minimum of the Madelung energy while $\zeta$ corresponds to the energy at $h=h_0$. Previously such investigations were performed by Nagai and Fukuyama in \cite{NF83} but for the Coulomb crystal only ($\kappa a=0$). They obtained $h_{\min}=1.633a_{\rm lat}$ and suggested that $h_{\min}=\sqrt{8/3}a_{\rm lat}$.

In addition to the hcp lattice, we considered another lattice with the hexagonal symmetry. We call it the ``MgB$_2$ lattice'' because it is the lattice of magnesium diboride under terrestrial conditions (space group P6/mmm). The MgB$_2$ lattice is a sequence of layers of magnesium and boron. The distance between adjacent layers is $h/2$, while $a_{\rm lat}$ is the distance between nearest magnesium ions in the layer. Number of ions in the elementary cell is $N_{\rm cell}=3$. Here we consider only the one-component Yukawa MgB$_2$ lattice formed by identical charged particles. Parameter $h$ is not fixed and is determined by the minimum of the Madelung energy. There is no experimental evidence that the MgB$_2$ lattice forms in Yukawa systems. However, this lattice possesses the forth smallest constant Madelung constant after the bcc, fcc, and hcp lattices (among known in the literature \cite{CF16,K18}). At $\kappa a=0$, the Madelung constant of the one-component MgB$_2$ lattice is equal to $-0.894505630008$. In the MgB$_2$ lattice $h_{\min}$ slightly depends on $\kappa a$ (see Tab. \ref{tab:1}). As in the hcp lattice, this dependence does not affect the computations noticeably and can be neglected. Further we consider that $h=\sqrt{8/3}a_{\rm lat}$ in the hcp lattice and $h\approx0.593936a_{\rm lat}$ in the MgB$_2$ lattice. Notice that the hexagonal lattice is not discussed because it is not stable in the harmonic lattice approximation \cite{K18} while possibility of formation this lattice was indicated in \cite{PG96}.

Madelung constants of the hcp and MgB$_2$ lattices are always larger than the Madelung constant of the fcc lattice (see Tab. \ref{tab:1}). At $\kappa a < 1.065714$ the bcc lattice has the lowest $U_{M}$ among all lattices in consideration while at $\kappa a > 1.065714$ the fcc lattice has the lowest $U_{M}$. This result agrees with Ref. \cite{FH94II}.

In \cite{FH94II} the electrostatic energy of the bcc Yukawa lattice was obtained by molecular dynamic simulations and fitted for $\kappa a < 1$ by a polynomial  (Eq. (15) from Ref. \cite{FH94II}). Our investigations allow to improve this approximation. Equation for the electrostatic energy at small $\kappa a$ can be obtained analytically from the expansion of Eq. (\ref{UTF}). It is clear to see from  Eq. (\ref{UTF}) that this approximation should contain only even powers of $\kappa a$. While fit from Ref. \cite{FH94II} keeps all powers.
\begin{eqnarray}
U^{\rm bcc}_{M}=&-&N\frac{Z^2e^2}{a}\left(0.8959292557+0.1037323337(\kappa a)^2\right. \nonumber \\
&-&0.0030913270(\kappa a)^4+0.0001430400(\kappa a)^6  \nonumber \\
&-&\left.7.1863 \times 10^{-6} (\kappa a)^8\right).
\label{Ubcc}
\end{eqnarray}
This equation represents $U_{M}$ with an accuracy of eight significant digit for $\kappa a<0.5$. The similar equation can be written for the fcc, hcp, and MgB$_2$ lattices:
\begin{eqnarray}
U_{M}=&-&N\frac{Z^2e^2}{a}\left(\zeta_0+\zeta_2(\kappa a)^2+\zeta_4(\kappa a)^4 \right. \nonumber \\
&+&\left.\zeta_6(\kappa a)^6+\zeta_8(\kappa a)^8\right), \label{UMg}
\end{eqnarray}
where parameters $\zeta_{i}$ are given in Tab. \ref{tab:fit1}.
\begin{table}
\caption{\label{tab:fit1} Parameters $\zeta_{i}$ for the Madelung energy of the fcc, hcp, and MgB$_2$ lattices.}
\begin{ruledtabular}
\begin{tabular}{cccc}
& fcc & hcp  & MgB$_2$  \\
\hline
$\zeta_0$ & 0.8958736152 & 0.8958381205 & 0.8945056294 \\
$\zeta_2$ & 0.1037956875 & 0.1040806163 & 0.1038098518  \\
$\zeta_4$ & $-0.0031060725$ & $-0.0031410914$ & $-0.0031091345$ \\
$\zeta_6$ & 0.0001451182 & 0.0001485484 & 0.0001455958 \\
$\zeta_8$ & $-7.4104\times 10^{-6}$ & $-7.4708\times 10^{-6}$ & $-7.7141\times 10^{-6}$ \\
\end{tabular}
\end{ruledtabular}
\end{table}

Equation (\ref{U}) gives the correct expression for the terms in the potential energy proportional to $(\kappa a)^0$ and $(\kappa a)^2$. Corrections to the charge density of the order of $(\kappa a)^4$ make same order changes to the potential energy.  At the same time the differences between the energies are extremely small and corrections of the order of $(\kappa a)^4$ and higher can radically change the structural diagram of the Yukawa crystals. The importance of these corrections is easy to illustrate by comparison of the two first terms in Eq. (\ref{UMg}). Let us denote the reduced electrostatic energy as
\begin{equation}
U_{M2}\equiv-N\frac{Z^2e^2}{a}\left(\zeta_0+\zeta_2(\kappa a)^2\right).
\end{equation}
Retention of only two first terms in $U_{M}$ leads to transformations of structural transitions. Indeed, at low $\kappa a$ the lowest $U_{M2}$ has the bcc lattice, at $0.93715<\kappa a< 1.58301$ the fcc lattice, at $1.58301<\kappa a<2.21838$ the hcp lattice, and at $\kappa a>2.21838$ the MgB$_2$ lattice.

\section{Zero-point energy}
Point-like charged particles in the crystal are not fixed and actually oscillate around their equilibrium positions even at $T=0$ due to the quantum zero point vibrations. Frequencies of these oscillations, $\omega_{\nu}$, can be found from the dispersion equation
\begin{equation}
\textrm{det}\{{D^{\alpha\beta}_{pp'}}\left(\textbf{k}\right)-
\omega^2_{\nu}(\textbf{k}) \delta^{\alpha\beta}
\delta_{pp'}\}=0, \label{Dis}
\end{equation}
where indices $\alpha$ and $\beta$ denote Cartesian components, $p$ and $p'$ run over the charged point-like particles in the elementary cell, $\nu$ enumerates the oscillation modes ($\nu=1,\dots,3N_{\rm cell}$) at a given wavevector $\textbf{k}$ and $D^{\alpha\beta}_{pp'}\left(\textbf{k}\right)$ is the dynamic matrix. The dynamic matrix of the Yukawa crystal with $N_{\rm cell}=1$ was derived in \cite{Ba02}. Equation for the dynamic matrix with arbitrary $N_{\rm cell}$ is given by Eq. (A1) from \cite{BK17} where the phonon properties were discussed in detail. It is instructive to compare the phonon spectrum obtained in \cite{BK17} with results of molecular-dynamic simulations directly and it will be done separately. Here we consider only the averaged over the volume phonon properties.

The dispersion equation (\ref{Dis}) allows to calculate the phonon spectrum at any wavevector $\textbf{k}$. Due to the periodicity of the crystal lattice it is sufficient to calculate $\omega_{\nu}(\textbf{k})$ only in the first Brillouin zone. Let us define the average of any function $f(\omega)$ of phonon frequencies over the volume of the first Brillouin zone $V_{BZ}=(2\pi)^3n/{N_\textrm{cell}}$
\begin{equation}
\langle f(\omega)\rangle= \frac{1}{3N_{\textrm{cell}}}\sum_{\nu=1}^{3N_{\rm cell}} \frac{1}{V_\textrm{BZ}}\int_\textrm{BZ} f(\omega_{\nu}\left(\textbf{k}\right))\textrm{d}\textbf{k}.
\end{equation}
Then the zero-point energy of the crystal is
\begin{equation}
E_0\equiv1.5N\hbar\langle \omega \rangle= 1.5N\hbar\omega_{p} u_1,
\end{equation}
where $u_1=\langle \omega/\omega_{p}\rangle$ is the first frequency moment, $\omega_{p} = \sqrt{4 \pi n Z^2 e^2/M}$ is the plasma frequency, $\hbar$ is the Plank constant.

\begin{figure}[ht]
\center{\includegraphics[width=1.0\linewidth]{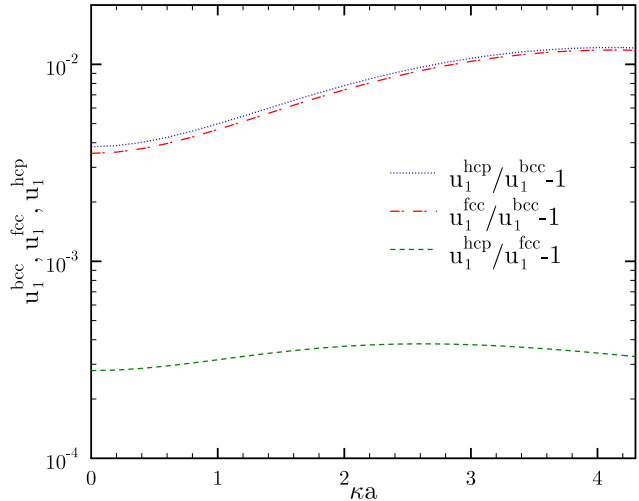}}
\caption{Ratios of lattice moments $u_1$ of the bcc, fcc, and hcp lattices.}
\label{fig:mom_pol_rac}
\end{figure}
In the Yukawa crystal, $u_1$ depends on the lattice type and $\kappa a$. In \cite{Ba02} the first moments of the bcc and fcc lattices were investigated and approximated for $\kappa a \ll 1$. The $u_1$ value of the hcp lattice is considered here for the first time. The one-component MgB$_2$ lattice is found to be unstable. At some $\bf k$, modes with $\omega_{\nu}^2(\textbf{k})<0$ appear in its phonon spectrum. Therefore the phonon properties of the one-component MgB$_2$ lattice were not considered. It is interesting to note that at $\kappa a=0$ the MgB$_2$ lattice with two different ions in the elementary cell (their charges are $Z_1$ and $Z_2\neq Z_1$) is stable at some values of $h$ and $Z_2/Z_1$. The bcc lattice is unstable at $\kappa a > 4.76$ in accordance with previous studies \cite{RKG88}. The fcc and hcp lattices are stable until $\kappa a = 5$ and we did not study these lattices at higher $\kappa a$.

Ratios $u_1^{\rm fcc}/u_1^{\rm bcc}-1$, $u_1^{\rm hcp}/u_1^{\rm bcc}-1$, and $u_1^{\rm hcp}/u_1^{\rm fcc}-1$ are plotted in Fig. \ref{fig:mom_pol_rac}. At $\kappa a=0$, $u_1^{\rm bcc}=0.5113877$, $u_1^{\rm fcc}=0.513194$, and $u_1^{\rm hcp}=0.5133369$. The relation $u_1^{\rm hcp}>u_1^{\rm fcc}>u_1^{\rm bcc}$ holds for any value of $\kappa a \geq 0$. In other words, the bcc lattice possesses the smallest zero-point energy at any $\kappa a$.

\begin{figure}[ht]
\center{\includegraphics[width=1.0\linewidth]{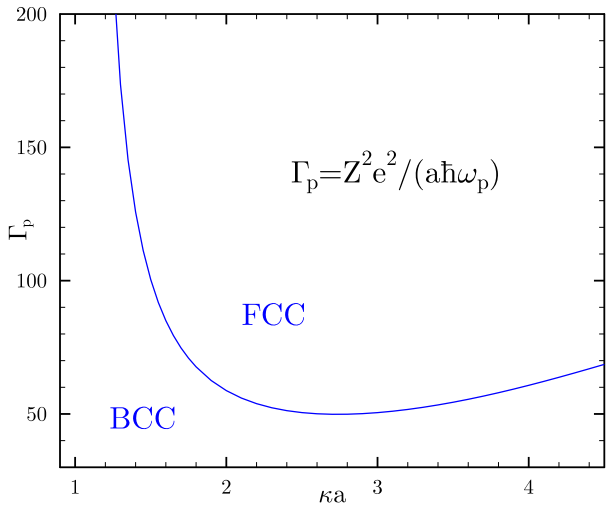}}
\caption{Dependence of $\Gamma_{p}$ on $\kappa a$ under condition $U_0^{\rm bcc}=U_0^{\rm fcc}$. Structural transition curve between the bcc and fcc Yukawa lattices at $T=0$ if the electrostatic and zero-point energies are taken into account.}
\label{fig:uf}
\end{figure}
At $T=0$, the total energy of the Yukawa crystal is
\begin{equation}
U_0\equiv U_{M}+E_0=NT_{p}\left(\Gamma_{p}\zeta+1.5 u_1\right),
\label{f_0}
\end{equation}
where $T_{p}=\hbar \omega_p$ is the plasma temperature. At $T=0$, the total energy is a function of two parameters, $\kappa a$ and $\Gamma_{p}\equiv Z^2 e^2/(a\hbar\omega_p)$. In Fig. \ref{fig:uf} we plot the dependence of $\Gamma_{p}$ on $\kappa a$ for which the total energy of the bcc lattice ($U_0^{\rm bcc}$) is equal to the total energy of the fcc lattice ($U_0^{\rm fcc}$). Above this curve, the energy of the fcc lattice is smaller than the energy of the fcc lattice. At $\Gamma_{p}\rightarrow \infty$ the energy of the zero-point vibrations can be neglected and $\kappa a$ tends to 1.066.

The dynamic matrix of the Yukawa crystals was derived from the same equation for the potential energy as the electrostatic energy, therefore $U_{M}$ and $E_0$ have similar precision. At the same time, $|u_1^{\rm fcc}-u_1^{\rm bcc}| \gg |\zeta^{\rm fcc}-\zeta^{\rm bcc}|$. Hence corrections to the zero-point energy $\propto (\kappa a)^4$ do not influence significantly to the transitions between lattices.

\section{Thermal contribution to the potential energy}
\begin{figure}[ht]
\center{\includegraphics[width=1.0\linewidth]{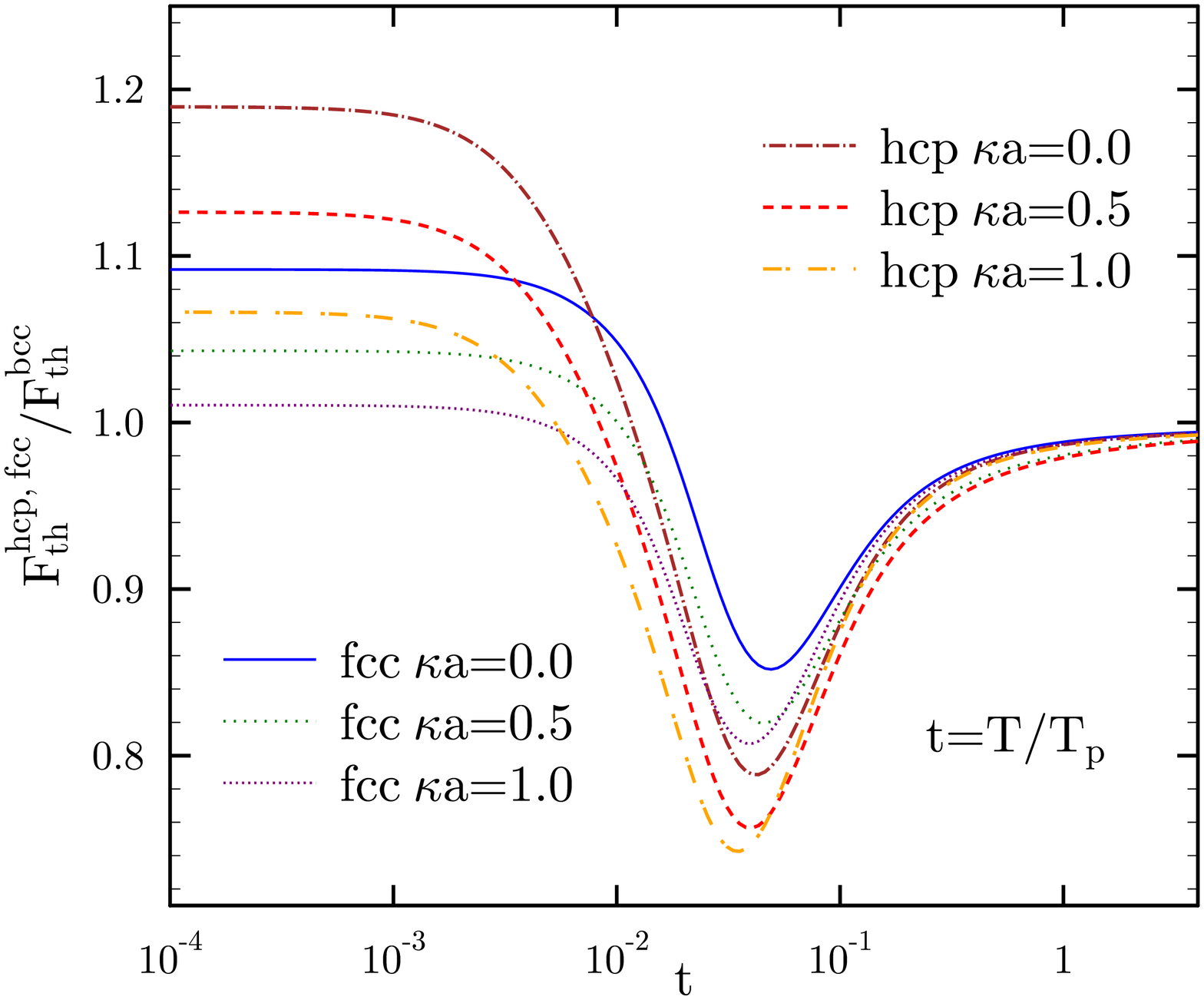}}
\caption{Ratios of $F_{\rm th}$ of the bcc, fcc, and hcp lattices.}
\label{fig:f_pol0}
\end{figure}

The phonon thermal contribution $F_{\rm th}$ to the total potential energy (thermal free energy) is equal to
\begin{equation}
F_{\rm th}=3NT\left\langle\ln\left(1-\textrm{e}^{-w}\right)\right\rangle,
\label{f_th}
\end{equation}
where $w\equiv\hbar\omega_{\nu}({\bf k})/T$ (e.g., \cite{BPY01}). For the bcc Yukawa lattice it was considered in \cite{Ba02}. Here we extend this study to other lattice types. In Fig. \ref{fig:f_pol0} we plot ratios $F_{\rm th}^{\rm fcc}/F_{\rm th}^{\rm bcc}$ and $F_{\rm th}^{\rm hcp}/F_{\rm th}^{\rm bcc}$ as functions of $t\equiv T/T_{p}$ for various values of $\kappa a$.

For crystals with the uniform background ($\kappa a=0$) these ratios were discussed in \cite{KB15}. At $\kappa a=1$ and $\kappa a=0.5$ they have the same features as at $\kappa a=0$. For instance, at high and medium temperatures ($T \gtrsim 10^{-2}T_{p}$) and any $\kappa a$, $F^{\rm hcp}_{\rm th}$ and $F^{\rm fcc}_{\rm th}$ are greater than $F^{\rm bcc}_{\rm th}$ ($F_{\rm th}$ is negative). At $\kappa a=1$, the ratio $F^{\rm hcp}_{\rm th}/F^{\rm bcc}_{\rm th}$ reaches minimum of $\approx 0.7425$ at $t\approx 0.03483$, while the minimum of $F^{\rm fcc}_{\rm th}/F^{\rm bcc}_{\rm th}$ is $\approx0.8072$ at $t\approx0.03926$. In the quantum limit, $F^{\rm hcp}_{\rm th}<F^{\rm fcc}_{\rm th}<F^{\rm bcc}_{\rm th}$ for any $\kappa a$. At $T \ll T_{p}$, the ratios $F^{\rm hcp}_{\rm th}/F^{\rm bcc}_{\rm th}$ and $F^{\rm hcp}_{\rm th}/F^{\rm fcc}_{\rm th}$ decrease with the increase of $\kappa a$. At $\kappa a=1$ they reach 1.066 and 1.0104, respectively. Figure \ref{fig:f_pol0} shows that phonon thermodynamic properties of different lattices at fixed $\kappa a$ and $t$ may vary from each other by tens of per cent.

\section{Total potential energy}
In the harmonic approximation the total potential energy $U$ (which better be called ``free energy'' but we use the same notation as in \cite{FHD97}) at any $T$ consists of three parts: the electrostatic (Madelung) energy $U_{M}$, the zero-point energy $E_0$ and the thermal contribution $F_{\rm th}$.  At $T \lesssim 5 \times 10^{-3}T_{p}$, the thermal contribution does not play a noticeable role and can be neglected. At these low temperatures it is enough to use Eq. (\ref{f_0}).

At high temperatures ($T \gg T_p$)
\begin{equation}
F_{\rm th}\approx3NT\left[u_{\ln}-\ln t\right]-1.5N \omega_{p} u_1,
\label{f_tht}
\end{equation}
where $u_{\ln}=\langle \ln{(\omega/\omega_{p})} \rangle$. At $\kappa a=0$, $u_{\ln}^{\rm bcc}=-0.831295$ and it is less than $u_{\ln}^{\rm fcc}=-0.8179085$ and $u_{\ln}^{\rm hcp}=-0.816031$. This situation holds at $\kappa a>0$. For instance, at $\kappa a=1$, $u_{\ln}^{\rm bcc}=-0.994814$, $u_{\ln}^{\rm fcc}=-0.978198$, and $u_{\ln}^{\rm hcp}=-0.976292$. At any $\kappa a$, the bcc lattice possesses the smallest $u_{\ln}$. The total potential energy $U$ is
\begin{equation}
\frac{U}{NT}=\Gamma \zeta+\frac{1.5 u_1}{t}+\frac{F_{\rm th}}{NT}=\Gamma \zeta+3\left[u_{\ln}-\ln t\right],
\label{F_ka}
\end{equation}
where $\Gamma\equiv Z^2 e^2/(aT)=\Gamma_{p}/t$ is the Coulomb coupling parameter. Hence, at high temperatures, the difference between the total potential energy of different lattices is independent of $t$ and can be considered as a function of $\Gamma$ and $\kappa a$ only.
\begin{figure}[ht]
\center{\includegraphics[width=1.0\linewidth]{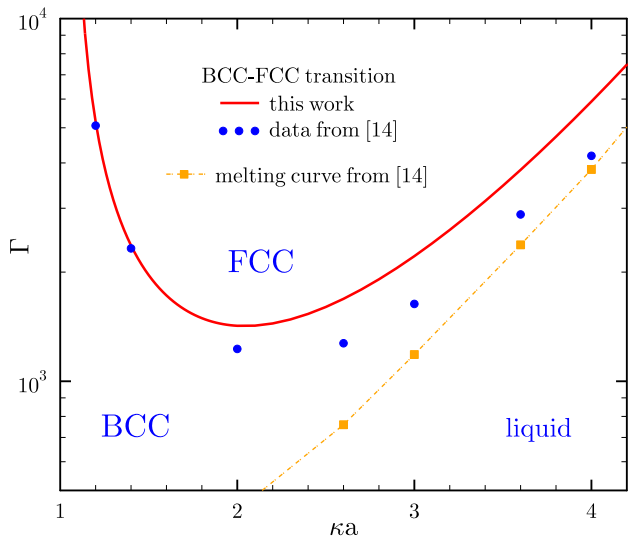}}
\caption{Phase diagram of Yukawa systems at $T \gg T_p$. Squares and circles are results of molecular-dynamic simulations from \cite{FHD97}, solid line is the result of the harmonic lattice approximation.}
\label{fig:f_pol}
\end{figure}

Structural transition curve between the bcc and fcc lattices at $T \gg T_{p}$ is plotted in Fig. \ref{fig:f_pol}. The solid line shows the result of our analytical calculations. It is similar to Fig. \ref{fig:uf}, but now the high-temperature limit is used. In the harmonic approximation $U^{\rm bcc}=U^{\rm fcc}$ at
\begin{equation}
\Gamma_{b}=3\frac{u_{\ln}^{\rm fcc}-u_{\ln}^{\rm bcc}}{\zeta^{\rm bcc}-\zeta^{\rm fcc}}.
\label{G_b}
\end{equation}
$\Gamma_{b}$ is given in Tab. \ref{tab:f_srav} for several $\kappa a$. Values of $\Gamma_{b}$ obtained from the molecular dynamic simulations in \cite{FHD97} are shown with points in Fig. \ref{fig:f_pol} and given in the $\Gamma_{b}^{\rm FHD}$ column in Tab. \ref{tab:f_srav}.

\begin{table}
\caption{\label{tab:f_srav} Values of $\Gamma_{b}^{\rm FHD}$ and $\Gamma_{b}$ for some $\kappa a$.}
\begin{ruledtabular}
\begin{tabular}{cccc}
$\kappa a$ & $\Gamma_{b}^{\rm FHD}$ & $\Gamma_{b}$ & $\Gamma_{b}^{\rm an}$  \\
\hline
1.2 & 5070 & 5201 & 5094 \\
1.4 & 2325 & 2369 & 2334 \\
2.0 & 1228 & 1422 & 1232 \\
2.6 & 1273 & 1688 & 1274 \\
3.6 & 2884 & 3827 & 2882 \\
\end{tabular}
\end{ruledtabular}
\end{table}
At $\kappa a < 2$ and far from the melting curve (dash-dotted line in Fig. \ref{fig:f_pol}), the difference between $\Gamma_{b}$ and $\Gamma_{b}^{\rm FHD}$ is small. At higher $\kappa a$ the discrepancy between our results and those of \cite{FHD97} can be explained by the absence of anharmonic corrections in our calculations. Analytically the first-order anharmonic correction to the energy of the Coulomb crystal was calculated in Ref. \cite{D90} and for crystals with $\kappa a >0$ have never been considered. In Ref. \cite{FHD97} corrections was approximated from numerical results as $A_1/\Gamma+A_2/\Gamma^2$. Notice, that $|A_1| \ll |A_2|$. If we add them to our calculations we receive a new $\Gamma_{b}$. In Tab. \ref{tab:f_srav} this new $\Gamma_{b}$ is marked as $\Gamma_{b}^{\rm an}$ and it agrees better with $\Gamma_{b}^{\rm FHD}$. The remaining differences, especially at $\kappa a=1.2$ and $\kappa a=1.4$, can be explained by the insufficient accuracy of previous computations. Near $\kappa a=1.066$, $\Gamma_{b}$ grows very fast, so high-precision calculations are needed.
\begin{figure}[ht]
\center{\includegraphics[width=1.0\linewidth]{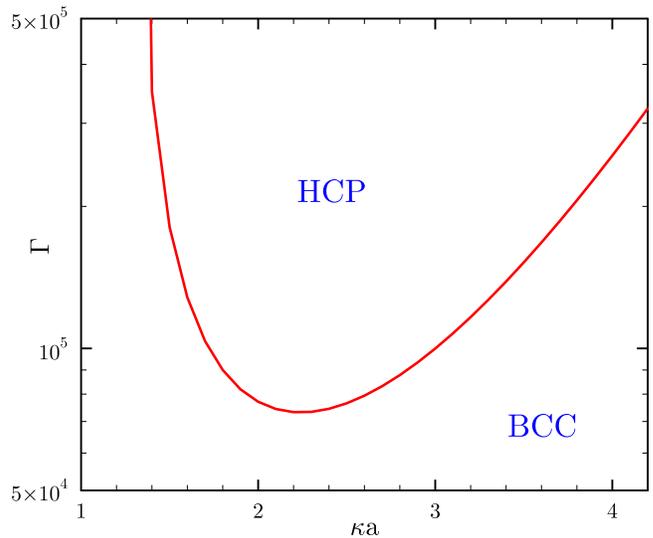}}
\caption{Structural transition curve between the bcc and hcp Yukawa lattices at $T \gg T_{p}$ in harmonic lattice approximation.}
\label{fig:f_pol_bh}
\end{figure}

The total energy of the hcp lattice is always higher than the total energy of the fcc lattice, but the transition between the bcc and hcp lattices can occur. It is plotted in Fig. \ref{fig:f_pol_bh}. Typical values of $\Gamma_{b}$ for this transition are order of magnitude higher that the typical values of $\Gamma_{b}$ for the bcc-fcc transition and for the phase transition between solid and liquid. $U^{\rm bcc}< U^{\rm hcp}$ at any $\Gamma$ if $\kappa a<1.30720$.

\begin{figure}[ht]
\center{\includegraphics[width=1.0\linewidth]{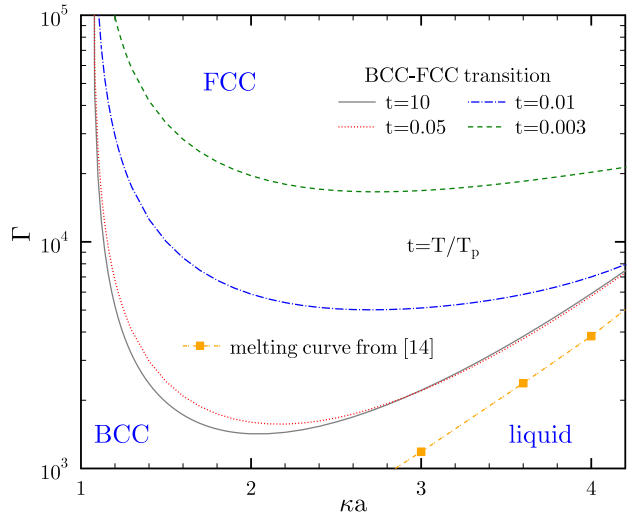}}
\caption{Phase diagram of Yukawa systems at different $T \lesssim T_{p}$. Squares are results of molecular-dynamic simulations from \cite{FHD97}, lines are results of the harmonic lattice approximation.}
\label{fig:f_polT}
\end{figure}

In contrast to classical molecular dynamic simulations, the harmonic approximation allows to calculate the total energy of Yukawa crystals at $T \lesssim T_{p}$. At such $T$, precise calculations of $F_{\rm th}$ should be used. According to Eq. (\ref{f_th}), the difference between the energies of the lattices is a function of $\kappa a$, $t$, and $\Gamma$. In Fig. \ref{fig:f_polT} we plot the dependence of $\Gamma_{b}$ on $\kappa a$ for $t=10$, $t=0.05$, $t=0.01$, and $t=0.003$. The solid curve for $t=10$ coincides with the solid curve in Fig. \ref{fig:f_pol}. A decrease of temperature leads to increase in $\Gamma_{b}$. At low temperatures, the thermal contribution can be neglected and $\Gamma_{b} \propto 1/t$.

\section{Conclusions}
The model of a crystal formed by point-like ions in the polarized electron background is widely used in the theory of neutron stars and white dwarfs (e.g., \cite{Ba02}). It turns out that this model is similar to the model of dusty strongly-coupled Yukawa crystal (e.g., \cite{FH94I}). It is shown that the electrostatic energy in both models is describing by the same analytical equation.

The electrostatic and thermodynamical properties of Yukawa crystals at $T \gg T_p$ were widely investigated in \cite{FH93,FH94I,FH94II,FHD97} by molecular dynamic simulations. In this paper we used the harmonic lattice approximation to calculate the total potential energy of such crystals and to verify results from Ref. \cite{FHD97} independently. In the harmonic approximation the total potential energy is a sum of electrostatic (Madelung), zero-point, and thermal free energies where the latter two contributions can be obtained from the phonon spectrum of the lattice. This approximation is successfully used to study the properties of crystals far from the melting point. Therefore at $\kappa a<2$ structural transition between the bcc and fcc lattices, which was obtained from molecular dynamic simulations in Ref. \cite{FHD97}, is analytically proved by our model (at $T \gg T_p$ structural transition between lattices depends on $\kappa a$ and $\Gamma$). At higher $\kappa a$ structural transition between the bcc and fcc lattices takes place near the melting curve (transition between the bcc lattice and the Yukawa liquid), where the anharmonic correct ions are needed to take into account.

Analytical harmonic calculations allow to consider other Yukawa lattices which have never been studied previously. In addition to the bcc and fcc we considered the hcp and MgB$_2$ lattices. It was shown that the MgB$_2$ lattice is unstable. While the total potential energy of the hcp lattice is always greater than the total potential energy of the fcc lattice so the new structural transition does not appear. Note that the difference between the energies of the bcc and fcc lattices is too small and next order corrections to the charge density can in principle lead to appearance the new transition.

Harmonic model allows to consider low-temperature effects which are difficult to examine by numerical simulations. At $T \lesssim T_p$ the total potential energy depends on $\kappa a$, $\Gamma$, and $t$.

\section{Acknowledgments}
The author is deeply grateful to P.S. Shternin for valuable comments to the paper. The work was partially supported by the Presidium of the Russian Academy of Sciences Program 13 ``Condensed matter and plasmas at high energy densities''.

\end{document}